\documentclass[manuscript]{aastex}
\slugcomment{Not to appear in Nonlearned J., 45.}

\shorttitle{Is there a hidden hole in Type Ia supernova remnants?}

\shortauthors{Garc\'\i a-Senz, Badenes, and Serichol}

\begin{document}


\title{Is there a hidden hole in Type Ia supernova remnants?}  


\author{D. Garc\'\i a-Senz\altaffilmark{1,2},   
C. Badenes\altaffilmark{3,4,5} and N. Serichol\altaffilmark{6}}
\altaffiltext{1}{Departament de F\'\i sica i Enginyeria Nuclear, UPC, 
Compte d'Urgell 187, 08036 Barcelona, Spain; domingo.garcia@upc.edu} 

\altaffiltext{2}{Institut d'Estudis Espacials de Catalunya, Gran Capit\`a 2-4, 
 08034 Barcelona, Spain}
\altaffiltext{3}{School of Physics and Astronomy, Tel-Aviv University, Tel-Aviv 69978, Israel; carles@astro.tau.ac.il}
\altaffiltext{4}{Benoziyo Center for Astrophysics, Weizmann Institure of Science, Rehovot 76100, Israel}
\altaffiltext{5}{Department of Physics and Astronomy, University of Pittsburgh, 3941 O'Hara Street, Pittsburgh, PA 15260, USA}
\altaffiltext{6}{Departament de Matem\`atica Aplicada III, Sor Eulalia d'Anzizu,
08034 Barcelona, Spain;  nuria.serichol@upc.edu}



\begin{abstract}

  In this paper we report on the bulk features of the hole carved by the companion star in the material ejected during a
  Type Ia supernova explosion. In particular we are interested in the  
long term evolution of the hole as well as in its fingerprint in the geometry 
of the supernova remnant after several centuries of evolution, which 
is a hot topic in current Type Ia supernovae studies.   
 We use an axisymmetric smoothed particle hydrodynamics code to characterize the geometric
  properties of the supernova remnant resulting from the interaction of this ejected material with the ambient
  medium. Our aim is to use supernova remnant observations to constrain the single degenerate scenario for Type Ia
  supernova progenitors. Our simulations show that the hole will remain open during centuries, although its partial or
  total closure at later times due to hydrodynamic instabilities is not excluded. Close to the edge of the hole, the
  Rayleigh-Taylor instability grows faster, leading to plumes that approach the edge of the forward shock.  We also
  discuss other geometrical properties of the simulations, like the evolution of the contact discontinuity.

\end{abstract}


\keywords{supernova: general, supernova remnants}


\section{Introduction}

The precise nature of the progenitors of Type Ia supernovae (SN Ia) is one of the most important unsolved problems in
modern astrophysics. Because of the key role played by SN Ia in the chemical and dynamical evolution of galaxies and in
cosmology the quest for these elusive progenitors has become a priority and a challenge to astronomers. Several lines
of observational evidence point to the thermonuclear explosion of a white dwarf as the most probable progenitor because
of the conspicuous absence of hydrogen emission lines in SN Ia spectra and the occurrence of these explosions in all
galaxy types. Theoretical considerations also favour white dwarfs as SN Ia progenitors, because triggering a
thermonuclear explosion in a degenerate object is not difficult, provided it has enough nuclear fuel to be burnt.
However, the puzzle starts just beyond this point because there are many different ways to explode a white dwarf, each
of them involving a different astrophysical scenario, \cite{bra95}, \cite{hill00}. 
 
One of these scenarios is the so-called single degenerate (SD) scenario, in which the explosion takes place in a compact
binary system composed by a massive white dwarf and a non-degenerate secondary star, either evolved or unevolved. If the
average mass transfer rate from the secondary is around $\simeq 10^{-7}$~M$_{\sun}$.yr$^{-1}$, surface ignition is
avoided and the white dwarf manages to gradually increase its mass, eventually approaching the Chandrasekhar mass limit \cite{hac96}.
At this point a carbon deflagration ensues close to the center of the white
dwarf. Conductive and convective heat transport mechanisms spread the combustion to the whole mass of the star in a time
scale of the order of a second, triggering a thermonuclear explosion. Hydrodynamic models of this kind of explosion
suggest that the observed nucleosynthetic yields and kinetic energies are better explained if the deflagration turns
into a detonation (i.e. supersonic burning) at some point, \cite{hoe96}, \cite{gam05},  
 but a consensus theoretical model for Type Ia supernovae does not exist yet.

One way to constrain the identity of SN Ia progenitors is to assume that the SD scenario is generally valid, and then
explore the observational consequences. Some of the implications of the SD hypothesis have to do with the effects that
the presence of the nearby secondary star has on the bulk properties of the supernova ejecta immediately after the
explosion, once the homologous stage has been reached. In this context, the presence of a companion star could modify
the picture of the explosion in several ways. a) if sufficient material is stripped from the outer layers of the
secondary during the collision with the ejecta, then this ablated, hydrogen rich, material should be seen in the SN
spectra, \cite{mbf00}. b) The large kick given to the secondary by the SN blast wave
would dramatically change its internal structure and deposit a large amount of linear momentum. In this case, the
observation of a peculiar star with large proper motion close to the expansion center of historical SN Ia remnants would
lend clear support to the SD scenario, \citep{can01}.  c) The presence of the secondary star can also
break the symmetry of the SN ejecta, introducing systematic effects that could be detected in spectral and
spectropolarimetric SN Ia observations, \cite{kas04}.

The first item above has been addressed in many works using both analytical calculations, \citet{wlmk75} and
hydrodynamic models in two \citep{mbf00} and three dimensions \citep{ser05,pak08}. All these calculations agree that, depending on the nature of the companion, 
the
stripped amount of hydrogen should range from  0.01-0.10 M$_{\sun}$~for main-sequence (MS) stars to $\simeq 0.5$~M$_{\sun}$~for red giants (RG), and therefore it should be detectable in SN Ia spectra
. Taken at face value, these calculations
would rule out the SD as one of the main channels to SN Ia because hydrogen has 
not been observed. Nevertheless, the issue is complicated because simulations
predict that a large fraction of the contaminating hydrogen should be moving at velocities below 1000 km.s$^{-1}$, where
it is mixed with Fe-peak elements that make its detection difficult because $H_\alpha$~ emission lines may blend with the many Fe and Co lines that appear 
during the nebular phase.
  On the other hand, observational constraints on the equivalent width of H lines in
SN Ia spectra measured during the nebular phase put rough limits on the maximum amount of low-velocity hydrogen
entrained in the SN ejecta. \citet{matt05} give an upper limit of 0.03 M$_\sun$ for SN 2001el, while \citet{leo07} quotes
upper limits of $0.01$~M$_\sun$~for SN 2005am and SN2005cf.
 
The unambiguous observational detection of the companion star of the white dwarf (item b) above) is the most direct and
effective way to constrain the progenitor scenario. A detailed search for such a star in the Tycho SNR was conducted by
\citet{prl04}, who claimed that a main sequence star with peculiar proper motion fulfilled all the criteria to be the
companion. Subsequent studies of the properties of this main sequence star have not been so conclusive
\citep{gh09,ken09}.

The imprint of the collision between the secondary star and the SN ejecta (item c) above) has also been analyzed in several
works. In particular, the hole carved in the otherwise spherical ejecta by
the shielding effect of the secondary is a source of asymmetry which could contribute to the observed diversity in SN
spectra. \citet{kas04} suggested a possible change of the peak magnitude with viewing angle of $\simeq 0.2$~mag in
$B$, comparable with the intrinsic dispersion of SN Ia light curves in this band. It was also found by the same authors
that the hole is a source of polarization in the observed spectrum. Even though polarization is in general low in SN Ia
spectra, it has been unambiguously detected and measured in surveys \citep[see][]{ww08}. Another observational signature
of the collision could appear hours or days after the explosion as a persistent optical/UV emission from viewing angles
looking down upon the shocked region, as suggested recently by \citet{kas10}. Nevertheless, a recent search for this emission, as reported by \cite{hay10}, did not give a positive result. 
All these
effects would be detectable during the SN phase. There are comparatively few studies on the late stages of the
evolution, once the ejecta start to interact with the ambient medium and the SNR phase begins. 
In a recent paper, \citet{vigh11} have analyzed the asymmetries introduced by holes
with varied apertures in the geometrical properties of Type Ia SNRs using multi-D hydrodynamics. They suggest that the
small asymmetries observed in the radius of the Tycho SNR can be interpreted as the imprint of the conical hole carved
in the ejecta at the moment of the collision. In order to explain these asymmetries, the authors had to assume a very
large angular width for the cone, $\simeq 90^0$. According to \citet{mbf00} such large aperture is only possible if the
secondary is a large red giant star, which would probably lose most of its weakly bound, H-rich envelope during the
collision with the ejecta. Such a large amount of stripped hydrogen $\simeq 0.4$~M$_\sun$~should be
visible in the spectra near maximum light and in the nebular phase, but this has not been reported in observations of
normal Type Ia SNe. In a recent paper \cite{lu11} have invoked the SD progenitor model 
for Type Ia supernova as the cause for the prominent  
X-ray arc clearly visible in the SW quadrant of the projected disc of Tycho SNR. The increase of the observed brightness at the X-ray arc zone was interpreted 
 by \cite{lu11} as the hallmark of the interaction between the supernova ejecta and the material stripped from 
the companion star shortly after the SN explosion. All this points    
to the SD scenario, in which the companion is an unevolved star, as a probable  
route to explain Type Ia SN explosions.  
   
The scarcity of studies of SNR evolution including the hole carved by secondary star at early times is the motivation
for this paper. Past studies have highlighted the great interest to carry out simulations to elucidate if the hole will remain open or not after several hundred years of evolution \citep{mbf00, pak08}. Our main goal is to follow the hydrodynamic evolution of the hole region as the remnant expands into an
homogeneous ambient medium (AM), and to infer possible observational consequences. 

The calculations presented below do not refer to a particular Type Ia SN remnant, rather they attempt to outline the gross features that the 
presence of a companion star imprints to the long term evolution of the SNR.   
The secondary was assumed to be a main sequence solar-like star, a possibility that has
been favored over red giant secondaries (see references in the discussion above). The choice of a companion mass of $1 M_{\sun}$~was also motivated because that is the case for which the detailed study of \citet{mbf00} provides more information (case termed HCV in their paper). In particular we want to check our 
estimation of the hole aperture and the velocity profile of the stripped matter inside the hole with that of Marietta and collaborators 
(described in their Sect.4.3). For example, they found that about half of the stripped material was confined in a cone within $\simeq 43^0$~from the symmetry axis which can be taken as a rough measure of the size of the hole. Using {\sl AxisSPH}~we have obtained that $\simeq 40\%$~of the stripped mass is within an angle of 
$40^0$. The slightly lower fraction of stripped mass within $\theta=40^0$~from our simulation may 
be due to the differences in the assumed explosion energy of the supernova ejecta ($\simeq 50\%$~ higher in Marietta et al.) as well as to the  different nature of the hydrodynamic schemes used to carry out the 
calculations.    
  Given the complexity of the calculations we have chosen to 
work with a unique scenario and to follow the interaction between the  
supernova ejecta and the companion star and later  
the surrounding AM. In this context, the choice of a solar-like star (a representative main sequence star) that is filling its Roche Lobe at the moment of the explosion is not unreasonable.

Simulations were carried out using an
axisymmetric SPH code developed recently, \citet{gs08}. Because it was not possible to simulate the evolution over the
large dynamic range in time between the collision and the fully developed SNR phase (minutes to thousands of years), we
divided the process in three stages. First, we studied the collision of the supernova ejecta with the secondary, spanning several
hours. After that time, the interaction stops and both the ejecta and stripped material are in homologous expansion. At this point the secondary was removed
from the calculation and the ejecta was stretched to a size of $\simeq 0.22$~pc, roughly the radius of the system 28 yr after the SN explosion. This radius is 
large enough for the ejecta to swept an appreciable amount of AM material.
 Finally, a large region of uniform AM was introduced around the SN ejecta, the
evolution of the SNR was followed until t$\simeq 1000$~yr, and the results were analyzed. 

This paper is organized as
follows. In Section 2, we briefly describe the main features of the hydrodynamic method we use and describe the initial
setting and the astrophysical scenario. Section 3 is devoted to the interaction between the supernova blast wave and the
main sequence star, and to a comparison between our results and those of other authors. In Section 4.1 we describe the
evolution of a spherically symmetric SNR as it propagates into the AM, as a benchmark to evaluate the asymmetric models
that include the hole. In Sections 4.2 and 4.3 we study the evolution of these asymmetric SNRs. We conclude with a final
discussion and a summary of our results.

\section{Hydrodynamic method and initial setting} 

We carried out the simulations with the axisymmetric SPH hydrocode {\sl AxisSPH}, described in \citet{gs08}, which
incorporates a new algorithm to solve the contribution of gravity in the axisymmetric SPH paradigm. The scheme
calculates gravity by computing the direct interaction between any pair of particles, each of them approximated as a
toroidal distribution of mass. To optimize computing time, the contribution of gravity was calculated only for the
particles belonging to the secondary star. Thus, the problem was modelled as a free supersonic fluid, -the SN ejecta-,
impacting on and passing through a gravitationally bound body, -the $1 M_{\sun}$~companion star-. This approach
precludes us from estimating the degree of contamination of the secondary by the supernova material, but it is a
reasonable approximation to model the structure of high velocity ejecta in homologous expansion.  During the collision
phase, the SN ejecta was represented by $10^5$~particles and the secondary star was simulated using
$2~10^4$~particles. The initial model for both components was obtained by distributing the mass particles uniformly in a
rectangular lattice and assigning them the mass required to reproduce the
spherically symmetric density profile of SN ejecta in the homologous phase and a polytropic solar-like star of
$1M_\sun$. The spherical model used for the supernova ejecta was the deflagration model by \citet{b96} with kinetic
energy at infinity of $8~10^{50}$~ergs.

The equation of state (EOS) used to study the interacion of the supernova ejecta with the companion star was that of an ideal gas plus radiation. Shock waves were handled by adding an artificial
viscosity term to the momentum and energy equations (see for example Rosswog 2009 for a recent review of SPH). In SPH,
the linear and quadratic terms of artificial viscosity are controlled by parameters $\alpha$~and $\beta$~respectively
which we set to $\alpha=1.5$~ and  $\beta=3$, slightly higher than the standard values to  
better handle with the hypersonic collision.  This particular setting helps to prevent artificial particle penetration between the SN ejecta and the AM during the first stages of the collision. The same values for $\alpha$~and $\beta$~ were chosen by \citet{hb92} to model the post-explosion hydrodynamics of 
SN 1987A with an axisymmetric SPH code, probably for the same reason. The center of the sun-like star was placed
$2,83R_\sun$~from the center of the explosion, at the point where the Roche-Lobe radius equals the radius of the secondary star. With that choice of parameters around 3\% of
the solid angle of the ejecta was intercepted by the secondary. This roughly corresponds to the HCV scenario considered
by Marietta and collaborators (2000).

Once the collision has come to an end, the companion star was removed from the computational box and the ejecta
structure was expanded homologously to a size of $\simeq 0.22$~pc. The ejecta was then surrounded by an uniform 
 AM with density $\rho_{AM}=3~10^{-24}$~g.cm$^{-3}$~and size 4 pc~$\times$~8 pc. A large number of
particles, $N_{AM}=500,000$, evenly spread in a rectangular lattice, were necessary to encompass this volume. The
structure of the ejecta was mapped with smaller number of particles $N_{ej}=8800$~in the innermost area, $0.22\times
0.44$~pc, of the computational domain. As gravity plays a negligible role in this phase 
the algorithm used to calculate gravity was
turned off. The EOS was changed to that of a perfect gas with 
$\gamma=5/3$.  
Once this basic spherically symmetric initial model was
built it was adequately modified to host a hole with the angular size estimated from simulations carried out with {\sl AxisSPH}~as described in the next section.

\section{Description of the interaction between the supernova ejecta and the secondary star}

In the absence of direct observations, the effects of the impact of the supernova ejecta on the companion star must be
addressed using numerical simulations. As stated above, the most complete numerical study to date is that of Marietta et
al. (2000), who used an axisymmetric hydrocode to follow the dynamical phases of the collision process and estimated
that $\simeq 0.1-0.2$~M$\sun$~were lost by a solar-like companion. The amount of stripped mass from the companion star
depends sensitively on the solid angle subtended by the secondary and to a lesser extent on the energy released by the
explosion. The general picture of the process obtained by Marietta et al. was later confirmed by \citet{ser05} and
\citet{pak08} using a three-dimensional SPH code. In the calculations of \citet{ser05}, the orbital movement of the
secondary was taken into account while it was not included in \citet{pak08} who conducted a resolution study of the event.
The incidence of the orbital dynamics in the outcome of the
collision is of minor importance for magnitudes like the mass stripped from the companion, the size of the hole carved
in the SN ejecta or the radial kick imparted to the companion star. However, other magnitudes are more sensitive to
the orbital dynamics, especially those related to the mixing of the stripped hydrogen and the chemically stratified
debris of the ejecta. In this case, the inclusion of the orbital velocity is crucial, because efficient mixing requires
that the different materials coincide not only in physical space but also in velocity space. A quantitative comparison
of several magnitudes of interest for the present work is given in Table 1.  
  The differences
in the stripped mass, the size of the hole, and the
velocity kick between the well resolved 2D model A and the low resolution 3D models  B and C of Table 1 are less than $20\%$, similar to the differences found by \citet{pak08}~in their resolution study. Some of
the discrepancies between the models shown in Table 1 come from the different prescriptions
used to model gravity in two and three dimensions.   

\begin{deluxetable}{lrrrrcrrrrr}
\tablewidth{40pc}
\tablecaption{Main features of simulated Type Ia supernova ejecta and $1M_\sun$~star }
\tablehead{
\colhead{Model}           & \colhead{M$_{sec}$}      &
\colhead{R$_\sun$}          & \colhead{D$_\sun$}  &
\colhead{V$_{ej}$}          & \colhead{M$_{inc}$}    &
\colhead{E$_k$/E$_b$}  & \colhead{~$\Delta$M}  &
\colhead{v$_{orb}^0$} & \colhead{v$_{kick}$} &
\colhead{$\Omega_{h}$}\cr
 &M$_{\sun}$& & &km.s$^{-1}$
&M$_{\sun}$& &M$_{\sun}$ &km.s$^{-1}$&
km.s$^{-1}$&deg}
\startdata
A (2D) & 1& 1& 2,83& 7500& 0,04 & 4,11 & 0,105 & \nodata & 86 & 40 \\
B (3D) & 1& 1 & 2,83& 7500& 0,04 & 4,11 & 0,09 & 232 & $72$ & 43 \\
C (3D) & 1& 1 & 2,83& 7500& 0,04 & 4,11 & 0,11 & \nodata & $96$ & 42 \\
\enddata
\tablecomments{When the orbit is included in the 3D calculation the width of the hole is slightly larger in the orbital
  plane $\Omega_{//}=43^0$~than in the ortogonal plane $\Omega{_\bot}=40^0$, \citet{ser05}. 
    }
\end{deluxetable}

\begin{figure}
\plotone{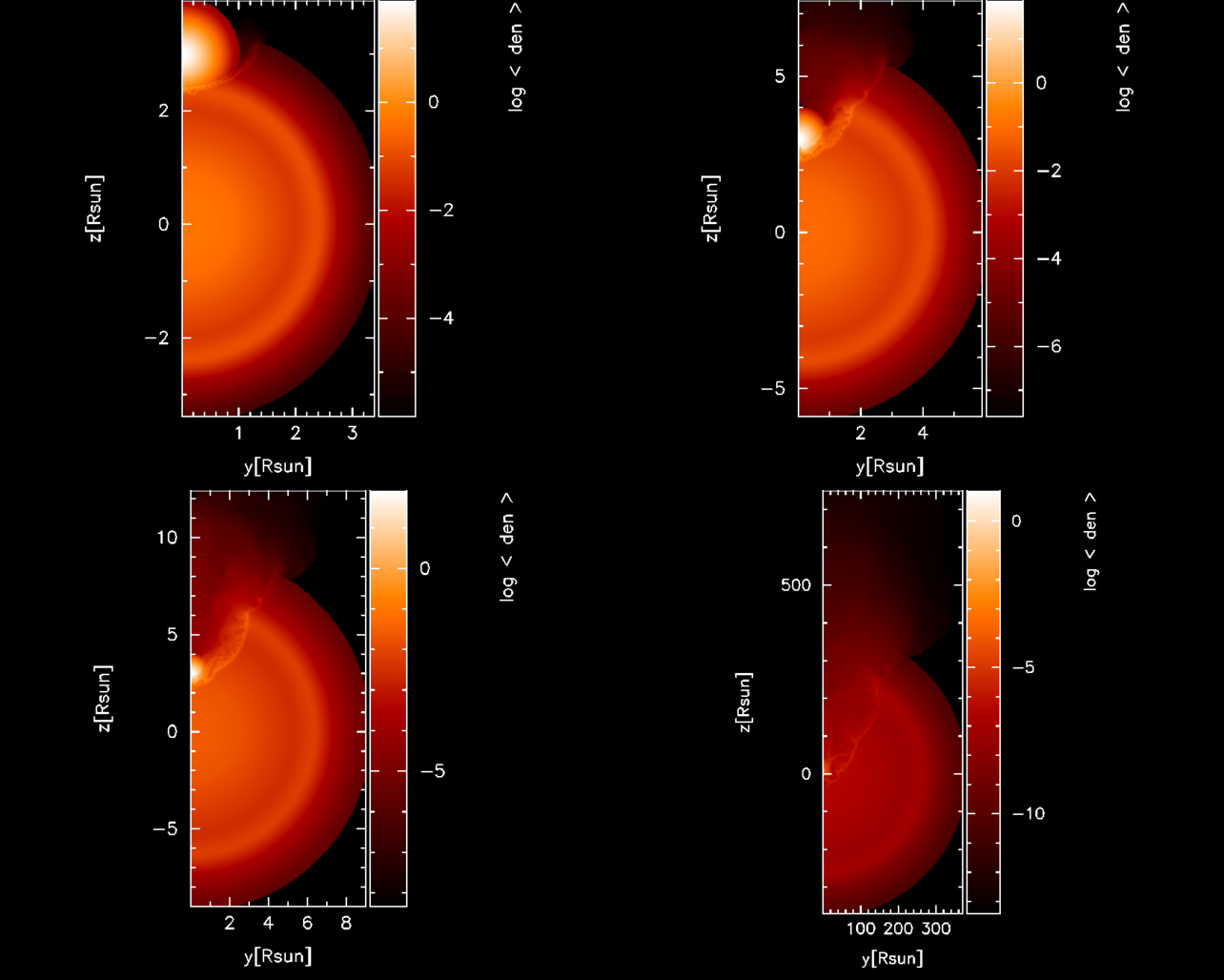}
\caption{Density snapshots summarizing the evolution of the collision of model A in Table 1. The first
  snapshot (top left) corresponds to $t=67 s$: the SN ejecta has compressed the frontal part of the envelope of the
  companion. At $t=190 s$~(second snapshot, top right) the SN ejecta has already wrapped around the companion. The third
  and fourth snapshots (bottom left and right) correspond to times $t= 340 s$~and t=17922 s.  The bow shock that is
  formed with the companion star at its apex can be clearly seen in all snapshots. The vertical dimension of the box
  changes from $2.2~10^{11}$~cm (first sanapshot) to 2.5~$10^{13}$~cm (fourth snapshot).  
    }
\label{fig1}
\end{figure}

The overall picture of the collision process can be seen in Figure 1, which shows density maps at different
representative times for model A in Table 1. The last snapshot corresponds to t$\simeq 5$~h after the
beginning of the collision. At this point, the interaction has ceased and the mass stripped from the secondary is $0.1
M_{\sun}$, in good agreement with models B and C, both calculated in 3D . The detailed temporal
evolution of the amount of stripped mass in the three models from Table 1 is shown in Figure 2. About half of the
stripped mass 
 is removed from the secondary in an initial violent episode lasting around 250 s, and the
remaining mass is removed more gradually during the interaction process. These two stages are usually referred to as the
stripping and ablation phases \citet{wlmk75}. Notice the brief episode of recapture of material which takes place
around  t= 210 s in the three models depicted in Figure 2. It is not clear whether this
feature, which is also present in the calculations of \citet{pak08}, is real or a numerical artifact due to the
simplicity of the criteria used to decide when a mass particle streams out from the secondary, namely that the actual
velocity of the stripped material exceeds the local escape velocity.  Such  
criteria neglect further hydrodynamic interactions of the stripped material 
being strictly valid only for material ablated just from the surface of the 
star.  
 
In Figure 3, we show the cumulative distribution of the SN ejecta and the stripped
mass in model A as a function of $\theta$, the angular distance to the axis defined by
the centers of the SN and the secondary star (henceforth we will use $\theta$ for
this angular distance, and $\theta_H=2~\theta$ for the complete opening angle of the hole).  
 In 
Figure 3 we can see how the {\sl hole} in the ejecta geometry caused by the shadowing effect of the
secondary is actually not empty, but filled with stripped material, basically H and He plus traces of heavier elements
. Nevertheless, the mass of SN ejecta inside the hole is
much lower than the mass stripped from the secondary. For angles smaller than $20^0$~the
region is almost devoid of SN debris, and it can still be called a {\sl hole}. The slope of the mass distribution
changes around $\theta=20^0$~(see Fig 3), suggesting that the aperture of the cone carved by the secondary is
$\simeq 40^0$, in good agreement to the calculations of Marietta et al. (2000) for a similar initial configuration. The
amount of stripped gas within an angle $\theta_H=40^0$~
is $\simeq 0.03 M\sun-0.04 M\sun$.

In Figure 4 (left) we show the radial velocity profile for the particles 
initially belonging to the secondary at t$\simeq 5$~h after the explosion. 
All that stripped material is moving faster than the approximate escape velocity from the secondary, which is represented by the horizontal line at 520 km.s$^{-1}$. As can be seen in Figure 4 (left) the stripped material is moving homologously, a feature which can be used to rescale the size of the system from hours after the explosion to years, when the interaction with the AM begins to 
affect the evolution of the remnant. In particular we have applied an homology 
 ratio of $5\times 10^4$~between the size of the ejecta at $t\simeq 5$~h, 
 the last calculated 
 model of the interaction with the companion star, and its  
size at $t\simeq 30$~yr, once  $\simeq 2~10^{-3}$~M$_\sun$~of AM have been swept by 
the SN ejecta.
  
We have already seen that a little less than a half of the stripped
material remains inside the hole created by the shadowing effect of the secondary, while the rest is outside of this
region, mixed with the SN ejecta. Past numerical simulations, \citet{mbf00}, \citet{pak08}, have shown that mixing
between stripped material and SN ejecta only affects low-velocity Fe-peak elements, because the other elements
synthesized in the SN are moving too fast to be mixed. The mass distribution in velocity space of the stripped material
inside the hole is shown on the right panel of Figure 4.  Most of the stripped mass is moving with velocities below 1000
km.s$^{-1}$,~with a peak in the distribution slightly higher than the escape velocity. 
  This is also in agreement with the results of \citet{mbf00}, who estimated a velocity of 823 km.s$^{-1}$~at the
half-mass point of the stripped material for a similar model.  The profile 
of mass distribution in velocity space depicted in Figure 4 has a characteristic high and low velocity tails connected by a sharp line with a peak around 
$\log~v(10^4~\mathrm {km.s}^{-1})=-1.3$. While the high velocity tail is probably well resolved by the simulations the low-velocity region is not 
so well represented due to the difficulties to set an exact criteria to decide when a particle moving close to the escape velocity becomes unbound (for example it could lose velocity after colliding and be recaptured). The mass distribution in velocity space of the stripped material given in Figure 4 will be used
in section 4 to set the initial model for the SNR evolution.

\begin{figure}
%\plotone{stripped.png}
\includegraphics[angle=270, scale=0.5]{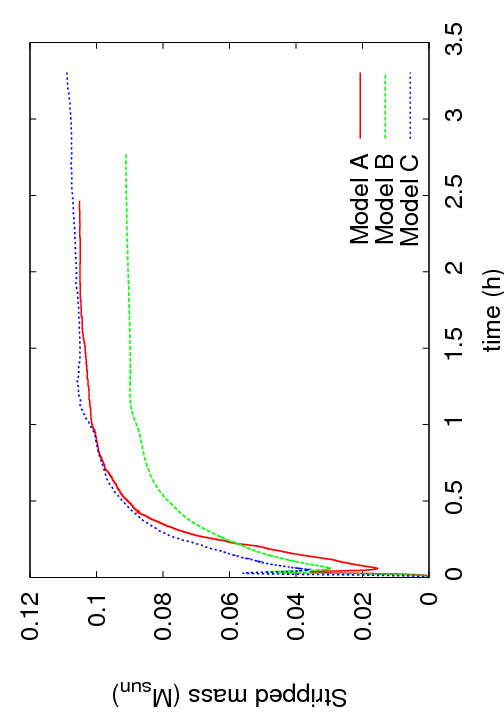}
\caption{Evolution of the stripped mass for models A, B and C of Table 1.}
\label{fig2}
\end{figure}

\begin{figure}
\includegraphics[angle=270, scale=0.5]{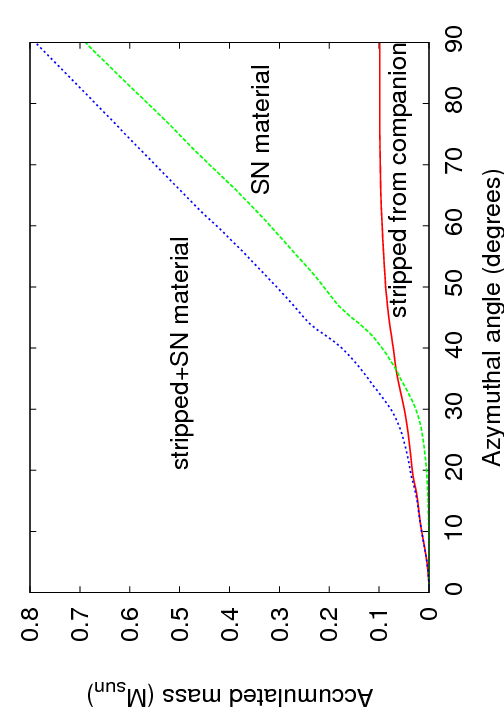}
%\plotone{f3.png}
\caption{Integrated distribution function of mass as a function of the azimuthal angle with the symmetry axis at
  t$=2840$~s for model A. The red solid line corresponds to the angular distribution of accumulated stripped mass. Long dashed line in green
  is for SN material while short dashed line in blue is for the whole sample of supernova material plus stripped mass. 
 For angles smaller than $20^0$~the region contains only
  material removed from the companion star.}
\label{fig3}
\end{figure}

\begin{figure}
%\plottwo{f4l.png}{f4r.png}
\includegraphics[angle=270, scale=0.6]{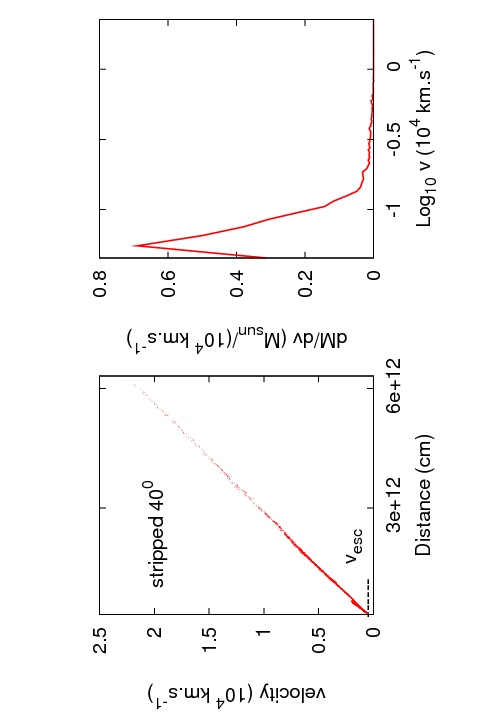}
\caption{Left: radial velocity distribution of the material stripped from the companion in a cone of
  $\theta=20^0$~at t$=17900$~s for model A. The short horizontal line at v$\simeq 520$~km.s$^{-1}$~is the approximate escape
  velocity for the surface of an unperturbed 1~M$_{\sun}$~polytrope. Right: Mass distribution of the stripped material
  in velocity space. 
 }
\label{fig4}
\end{figure}

\section{Interaction with the ambient medium}

\subsection{SNR models with spherical ejecta}

The large scale structure of SNRs is determined by the interaction between the SN ejecta and the surrounding ambient
medium (AM).  In this work, we consider a uniform AM with a density $\rho_{AM}= 3~10^{-24}$~g.cm$^{-3}$. That choice is supported by \cite{bad06} and \cite{bad08}  
who have shown that the fundamental dynamical
  and spectral properties of Type Ia SNRs can be reproduced by models that assume a uniform AM. Also \cite{bad07}
  showed that other kinds of AM (in particular, large wind-blown bubbles excavated by the accretion winds predicted by
  some Type Ia SN progenitor models) are inconsistent with the sizes of historical Type Ia SNRs. The adopted value for the density  
  is representative for the warm phase of the ISM in our galaxy, \cite{fer01}, 
	and close to the best fit value for Tycho. The typical size of the SNR
at the
time that a significant mass fraction $f=\frac{\delta M_{AM}}{M_{SN}}$~has been swept by the blast wave is:

\begin{equation}
R=\left(\frac{3~f~M_{SN}}{4\pi\rho_{AM}}\right)^{\frac{1}{3}}
\end{equation}

\noindent
which gives radii of $\simeq 0.5$~pc, 1.08 pc and 2.3 pc for f=0.01, 0.1, and 1, respectively.  We have started the simulations of the
SNR stage when the size of the SNR is $R_0=0.22$~pc, corresponding to $\simeq 28$~years after the SN explosion, to
ensure that the swept-up mass is still very small. Such initial radii roughly corresponds to the size of the SN ejecta at the time when its outer density just equals the assumed density of the AM, $\rho_{AM}=3~10^{-24}$~g.cm$^{-3}$. It corresponds to a value of f$\simeq 0.1\%$~in Equation (1). The starting time, 
t$=28$~yr is, however,  larger than 
that considered in the 2D calculation of \cite{dwa00} who  made use of an expanding moving grid 
to increase the resolution during the very first stages of the SNR. As far as the initial swept-up mass is small enough the evolution of the 
remnant does not depend too much on the exact location of the contact discontinuity at the initial simulation time, \cite{dwa98}.    
Taking
$f=1$~in Equation (1) we can obtain the size of the SNR when it enters the Sedov
  stage: R$\simeq 2$~pc. At that moment the age of the SNR would be a few hundred years.

\begin{deluxetable}{lrrrcrrrr}
\tablewidth{30pc}
\tablecaption{Features of SNR models at t'=0.11}
\tablehead{
\colhead{Model} &\colhead{Geometry} &
\colhead{Composition ($\theta\leq 20^0$})& $\rho_{AM}$~(g.cm$^{-3}$)}
\startdata
D & Spheric&  Type Ia-SN & $3\times 10^{-24}$ \\
E & Empty hole&  \nodata  & $3\times 10^{-24}$\\
F & Filled hole&  solar & $3\times 10^{-24}$ \\
\enddata
\end{deluxetable}

In order to benchmark the results obtained with the ejecta models modified by the collision with the secondary, we have
also simulated the evolution of a SNR with spherical ejecta, quoted as model 
D~in Table 2. We expanded the SN ejecta from its initial size to a SNR
age of t= 28 yr (outer radius $\simeq$~0.22 pc). The density and velocity profiles after this homologous expansion are
shown in Figure 5. At this age, the ejecta only fills a small fraction of the computational space, which, even in 2D, is
a challenge for the numerical simulation. To build the initial model we have followed the method described in
\citet{vel06}. The SN ejecta was represented by a small subset of N$_{SN}=8114$~particles, evenly spread in a
rectangular grid of $(0.22\times 0.44)$~pc$^2$, while the AM was represented with N$_{AM}=562,086$~particles spread in
a regular lattice of $(4\times 8)$~pc$^{2}$. The mass of the ejecta particles was adjusted to reproduce the density
profile given in Figure 5. Although the total number of particles in the ejecta is small, the resolution is sufficient
to follow the main features of the SNR evolution, including the development of the supersonic forward and reverse shocks
\citep{vel06}. 

 The use of particles with different mass to reproduce the initial density profile warrants clarification because it could be the source of numerical artifacts which may distort the simulation. According to Figure 5 it is evident that the mass of the particles must change in a factor of the order of $10^3$~in the neighborhood of the contact surface. Such large mass ratio 
may be a  
source of numerical troubles when the supernova particles and those of the AM mix. However, the real situation is not as bad as it may seem owing to the peculiarities of the mass of the particles in axisymmetric SPH codes. In the axisymmetric geometry the local average of a physical magnitude $A$~is calculated as 
$ <A>_i =\sum_j \frac{m_j}{2\pi r_j \rho_j} A_j W_{ij}$~(where $r_j$~is the distance 
of the particle to the Z-axis) from which $<\nabla A>_i$~is 
conveniently estimated, \citet{gs08}. Therefore particles  carry   
 an effective mass $m'=\frac{m}{2\pi r}$~to weight the interpolations. When the supernova material expands such effective mass is reduced. After one hundred years, when material begins to be mixed with the AM, the outer edge of the ejecta has already changed its radius by a factor $\simeq 5$. When the ejecta reaches the limits of the system, that factor has grown to $\simeq 20$. Due to the RT instability the supernova matter floats inside the AM whereas the less dense AM material falls down so that the initial contrast in effective masses is progressively reduced as the 
SNR evolves.

 Probably main effect of working with particles of different masses is that they increase the numerical noise which in turn serve as a seed of the RT instability during the first stages of the evolution. 
A similar conclusion was reached by \cite{hb92} and \cite{hw94} who used an axisymmetric SPH code to simulate the post-explosion hydrodynamics of SN 1987A. The density snapshots depicting the evolution of the RT instability in these works are asthonishingly similar to these of ours, especially in the Herant \& Woosley simulations (for instance compare their 
Figure 3 with our Figure 6 below). What is reassuring is that they did also use particles with different mass but a quite different geometry for the initial lattice. 

To check more quantitatively the impact of using particles with different 
masses we ran a calculation in which the number of particles belonging to the ejecta was doubled (and their mass consequently halved). A second simulation was launched by doubling the number of particles everywhere so that resolution was a 
  factor $\sqrt{2}$~better. We have no detected significant differences in the evolution of the SNR. Apparently the main effect of doubling the number of particles in the SN region is to slightly delay the time at which the RT instability appears. Thus the impact of using particles with variable mass in the 
simulations seems to be limited. 
It is a source of numerical noise that may distort the growth of the Kelvin-Helmholtz instability during the process of mixing, but we do not think it is seriously affecting the large scale features of the RT instability. It could also contribute to imprint the more filamentary look on the RT mushroom caps seen in the simulations presented below. 

The deceleration of the ejecta caused by the AM is equivalent to a local gravitational field pointing toward the center of the
explosion, favoring the growth of the Rayleigh-Taylor (RT) instability in the dense layer between the forward and the
reverse shocks, \citet{che92}.  
In our simulations, the growth of the RT instability was induced by numerical noise in the initial
distribution of the particles. As
we will see, the development of the RT instability has a large impact on the long term evolution of the hole.  Several
snapshots of the density evolution of the SNR for model D are shown in Figure 6 where times were given in normalized units $t'= t/T$~being T defined as 
in \cite{dwa00}:
 
\begin{equation}
T= 248~E_{51}^{-0.5}\left(\frac{M_{ej}}{M_{Ch}}\right)^{5/6}\left(\frac{\rho}{
2.34~10^{-24}~\mathrm g.cm^{-3}}\right)^{-1/3}~\mathrm {yr}
\end{equation}
 
\noindent which for our parameter choice becomes T=250 yr.
In the first snapshot, $t'=0.43$~after the supernova explosion 
 (t'$_{sim}$=0.32, henceforth we refer as t' and t'$_{sim}$~the normalized 
elapsed times 
since the explosion and from the beginning of the simulation respectively),  
the shocked ejecta still retain spherical
symmetry.  
In the second snapshot, t'= 0.87, the
RT instability has grown significantly, and the shocked ejecta begins to fill the interaction region between the reverse
and forward shocks. The instability develops and gains complexity in the third and fourth images
. In the last picture,
the forward shock has already reached the limits of the computational domain, and the reverse shock, although still far from the explosion site, has gone back 
through more than the 95\% of the supernova material. 
 The large plumes that develop at late times close to the symmetry axis are numerical
artifacts due to the imposed axial geometry. The mass inside these plumes is very small, and it does not affect the
outcome of the simulations in a significant way. The left panel in Figure 7 summarizes the evolution of the spherical
SNR at the moment when the blast wave have reached the limits of the computational domain.  

The onset of the self-similar Sedov stage depends on the AM density, kinetic energy of the ejecta and also on the precise profile adopted for the ejecta, \cite{dwa98}. For the 
adopted values of these magnitudes in model  D the swept-up mass becomes comparable to the ejecta 
mass at t'$\simeq 0.92$. Two dimensional simulations of SNR were carried out by \cite{che92} and \cite{dwa00}. In particular \cite{dwa00} investigated in some detail the interaction between the supernova ejecta and the ambient medium, either assuming constant AM as well as a CSM whose density decreases as $r^{-2}$. These calculations were carried out using the finite-difference code VH-1 an assuming an exponential density profile for the ejecta. As our calculations were done 
using a SPH code with a different ejecta profile (see left panel in Figure 5) 
and a larger $\rho_{AM}$ than in \cite{dwa00} a brief description of the evolution is warranted. During the self-similar evolutionary stage the average contact discontinuity radius evolves $R_{CD}\propto t^s$~where the mean value for the 
exponent deduced from our simulations, $s=0.42$~(see left panel in Figure 7) agrees with the theoretical value $s=0.4$~expected for the Sedov solution. In the 
work of \cite{dwa00} the expansion parameter approach the analytical value after a time t'$\simeq 3-4$.  
The normalized times used to find 
the expansion parameter in Figure 7 go from $t'= 0.9$~to $t'=3.6$~hence 
our results roughly agree to that of \cite{dwa00} but convergence to the 
theoretical value is a little faster. Such small discrepancy may arise from the 
differences in the ejecta profiles used in both simulations.
There also differences in the 
morphology of the RT fingers, more filamentary in the SPH calculation,  
which extends to a larger radius in the unstable layer. On the other hand the initial phase of growth is is slower in the SPH calculation owing to the lower resolution (in the VH-1 simulation a moving grid with a 
constant number of computational cells was used) and to the damping introduced by the artificial viscosity.   
 
The trajectory of the contact discontinuity is shown in the right panel of Figure
7, and it can be roughly fitted by a parabola, $R_{CD}=0.5~a_{CD}~t'{^2}$~assuming a deceleration $a_{CD}=0.2~pc$~in adimensional time units. 
. The growth rate $\Gamma$~of the
RT instability during the linear regime can be estimated using:

\begin{equation}
\Gamma=\sqrt{\frac{A_t~n~a_{CD}}{R_{CD}}}
\end{equation}

\noindent where $A_t$~is the Atwood number across the interface and $n$ is the wave number 
. Once the swept-up mass has become comparable to the ejecta mass, at $t'\simeq 0.92$~and R$_{CD}\simeq 2$~
pc, we get $\Gamma=0.22\sqrt{n}$~ for $A_t=0.5$.  After $\Delta t'= 1$, the e-folding growth factor,
$e_{gf}$~ of a small perturbation is $e_{gf}\simeq 0.22\sqrt{n}$. Therefore, only perturbations with high
wave  number are in the non-linear regime at this time. However, in SPH small perturbations are usually damped by the
artificial viscosity, so we expect that the RT instability will remain in the linear regime except at late
times when the forward shock approaches the limits of the system.  
     
The location  of the contact discontinuity follows
the $R_{CD}(t')$~average trajectory, modulated by the effect of the RT instability.  In Figure 8, we
show the location of the contact discontinuity as a function of $\theta$.  To obtain the location of the CD, we
recorded the position of the ejecta particle with the largest radius at each azimuthal angle.  
 The fundamental features of this profile can be compared to those of the profile measured by \citet{warren05}
for the CD in Tycho's remnant. The comparison, however, is not completely straightforward, because the profile observed
by Warren et al. results from projecting the CD surface onto the plane of the sky, while our simulations provide a
section of this surface, relative to the explosion center, without projection 
effects. Another caveat stems from the axisymmetric hypothesis, which
constrains the growth of the RT structures to have axial, ring-like, symmetry. 
In this respect \cite{blo01} did not find large differences between two and 
three dimensional simulations for an EOS with $\gamma=5/3$. There is an increased amount of small scale 
structure in the 3D calculation as well as a slightly larger penetration of the RT fingers but the amount of turbulent energy in the unstable layer was similar. It is also worth to comment that for $\gamma < 5/3$~(as it would be the 
case if there is particle acceleration at the shock front) the differences between 
2D and 3D cases are larger, \cite{blo01}.   
 
The power spectrum (PWS) of the azimuthal distribution of the radial amplitude
fluctuations of the contact discontinuity depicted in Figure 8 is shown in 
the leftmost panel of Figure 9. 
In the following section, we will compare these PWS to those resulting from ejecta models that include a
hole. Both PWS shown in Figure 9 correspond to the last snapshot in Figure 6, but are calculated using only
the upper ($\theta\le 90^0$) and the bottom quadrants ($90^0 <\theta\le 180^0$) 
of the computational space respectively. The PWS that we obtain is much flatter than in
\citet{warren05} (see their Figure 6), and it shows larger fluctuations. 
 The mode with the largest power corresponds to n=1 (or a wavelength in the azimuthal direction of 90$^0$),
which probably comes from the imposed symmetry between quadrants at t'$_{sim}$=0.  As expected, the PWS in different quadrants
of the spherical ejecta simulation are very similar to each other. A linear fit to both spectra gives $P_{w}\propto
k^n$~with $n\simeq -0.5$~, much lower than $n\simeq -1.5$~obtained by \citet{warren05} for Tycho, although we remind the
reader that this comparison is not straightforward. 
 
\begin{figure}
%\plotone{f5.png}
\includegraphics[angle=270, scale=0.6]{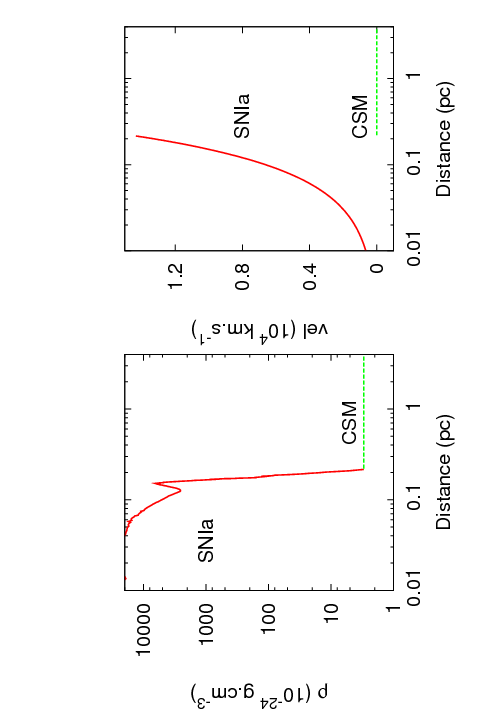}
\caption{Spherically symmetric density and velocity profiles of SN Ia ejecta at the time the simulation of the
  interaction with the AM starts. }
\label{fig5}
\end{figure}
      
\begin{figure}
\plotone{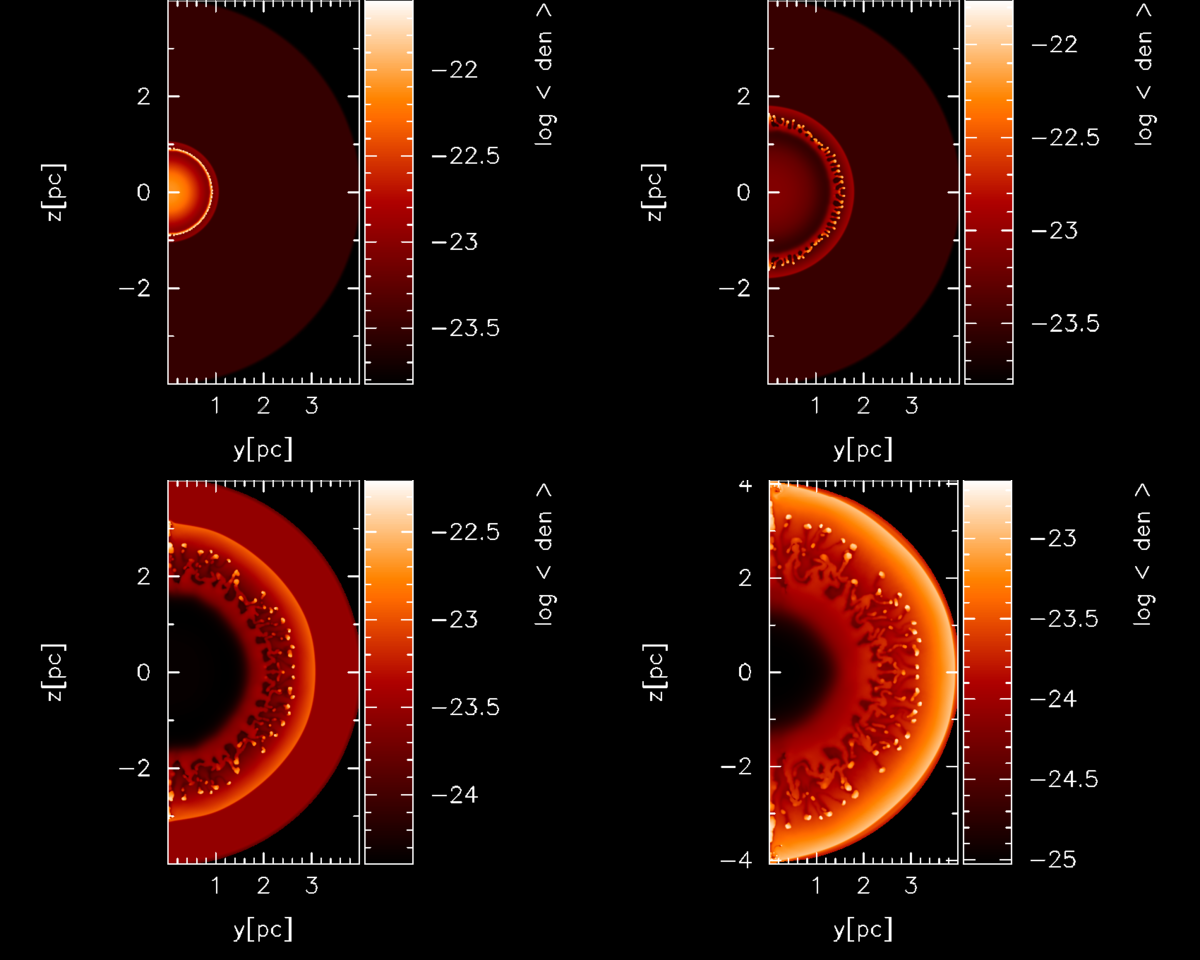}
\caption{Snapshots summarizing the evolution of the density during  
the Sedov-Taylor self-similar phase of the  
SNR corresponding to model D in Table 2. Adimensional elapsed times from the 
supernova explosion are t'=0.43, t'=0.87, t'=2.2 and t'=3.6. 
}
\label{fig6}
\end{figure}

\begin{figure}
%\plotone{f7.png}
\includegraphics[angle=270, scale=0.6]{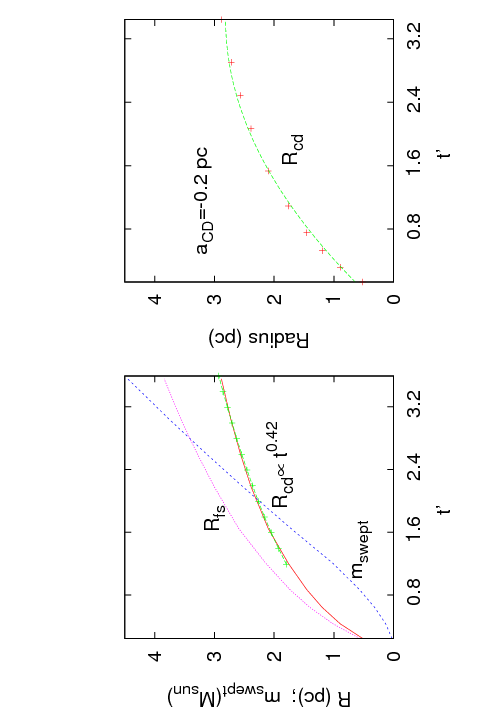}
\caption{Main features of the evolution of model D (left): the swept-up mass of the 
AM, the average radius of the contact discontinuity and the radius of  
the forward shock. A parabolic fitting for the locus of the CD
is provided in the rightmost figure. The deceleration of the contact surface is a$_{CD}\simeq 0.2$~pc in adimensional time units. 
}
\label{fig7}
\end{figure}

\begin{figure}
%\plotone{f8.png}
\includegraphics[angle=270, scale=0.6]{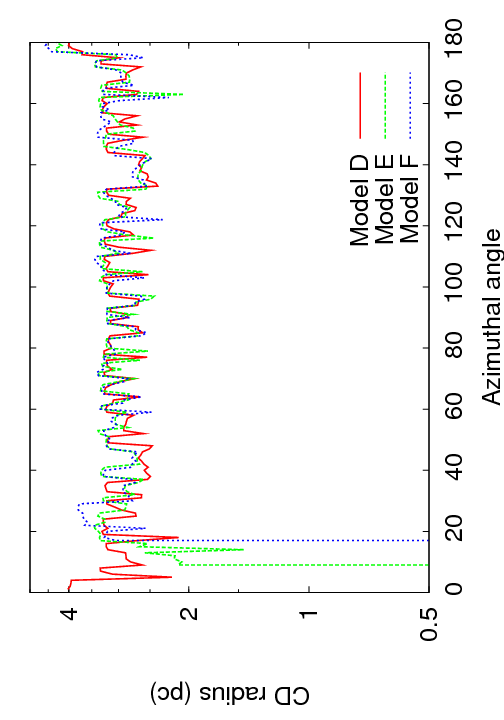}
\caption{Radius of the CD as a function of the azimuthal angle $\theta$~for models D, E and F of Table 2 at adimensional times t'=3.6, 3.9 and 3.9 
respectively.} 

\label{fig8}
\end{figure}

\begin{figure}
%\plotone{f9.png}
\includegraphics[angle=270, scale=0.7]{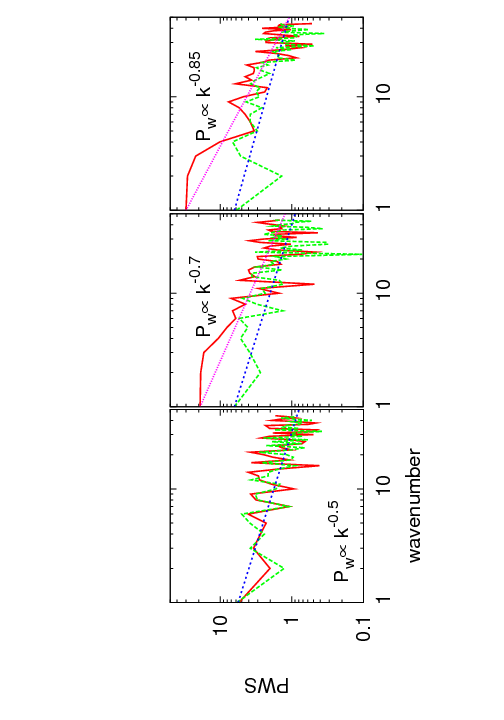}
\caption{Power spectra of the CD locus shown in Figure 8 for models D (left), E (center) and F (right). Continuum lines are for the upper quadrant hosting the hole 
whereas dashed lines are for the lower quadrant. Fitting straight lines for 
the three models are also depicted. 
}
\label{fig9}
\end{figure}

\begin{figure}
\plotone{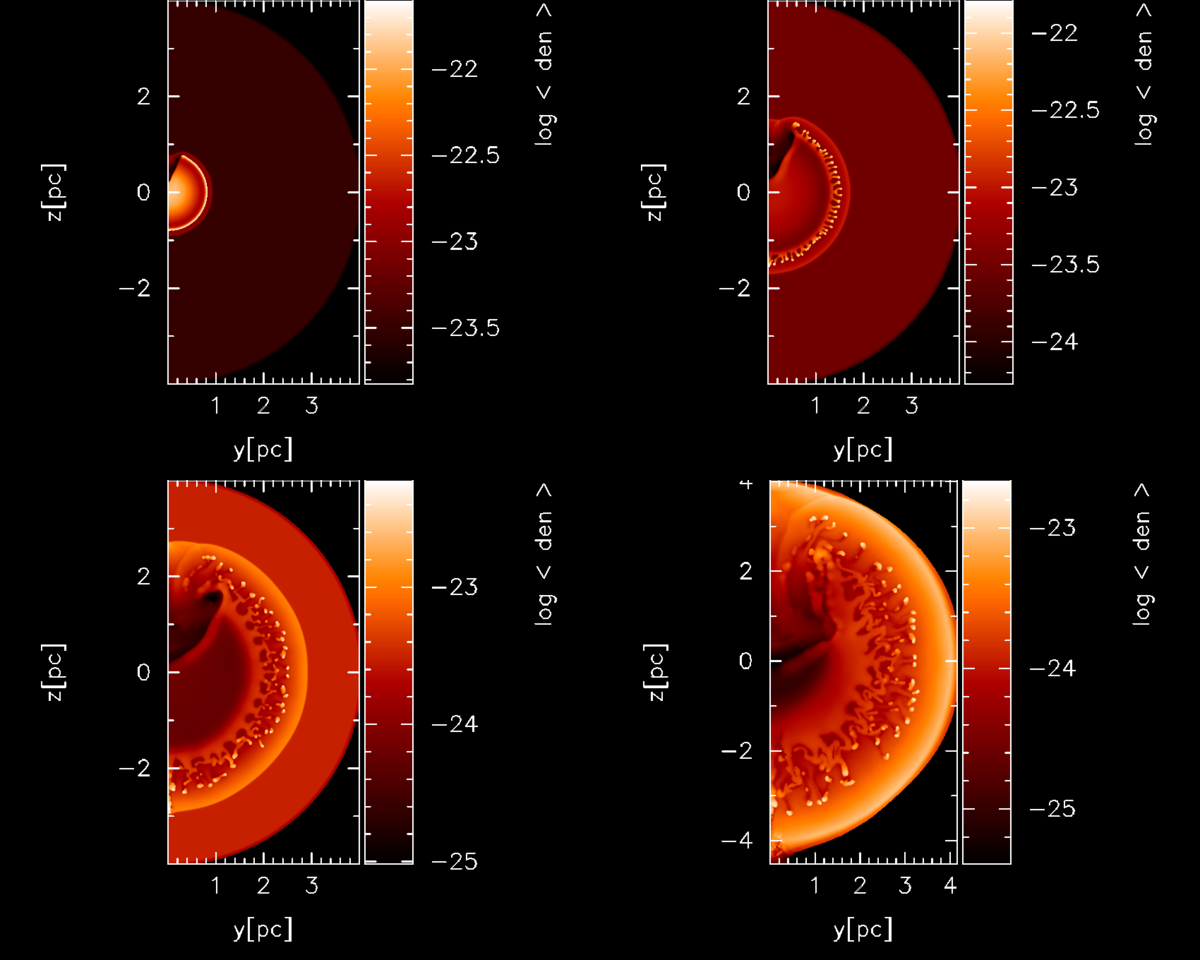}
\caption{Snapshots summarizing the evolution of the density during
the Sedov-Taylor self-similar phase of the
SNR once a $\theta=20^0$~cone of material belonging to the ejecta was removed from the
 simulation, model E of Table 2. Adimensional times are t'=0.38, t=0.81, t'=1.95 and t'=3.4 from the 
beginning
of the simulation.
}
\label{fig10}
\end{figure}

\begin{figure}
%\plotone{f11.png}
\includegraphics[angle=270, scale=0.65]{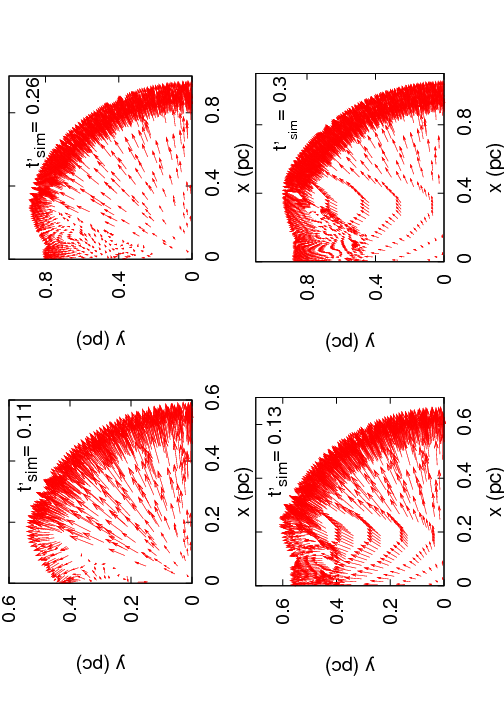}
\caption{Velocity map around the hole for models E (upper row) and F (lower row) at early times.   
}
\label{fig11}
\end{figure}

\begin{figure}
%\plotone{f12.png}
\includegraphics[angle=270, scale=0.65]{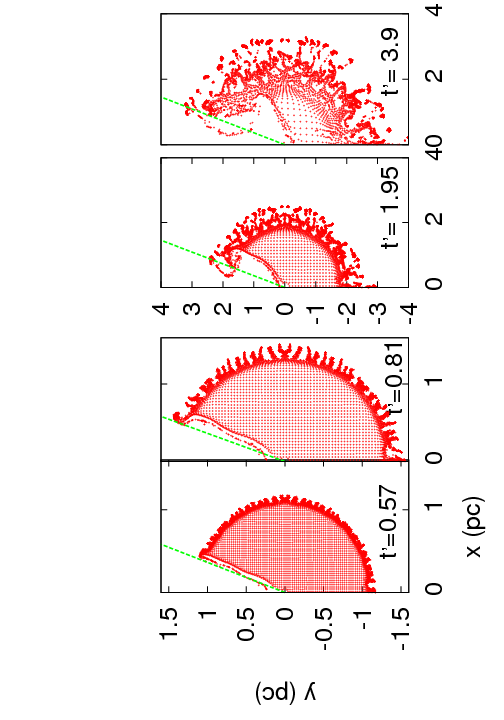}
\caption{Intrusion of the supernova material toward the symmetry axis impelled by 
the RT instability (model E of Table 2).
}
\label{fig12}
\end{figure}

\begin{figure}
\includegraphics[angle=0, scale=0.5]{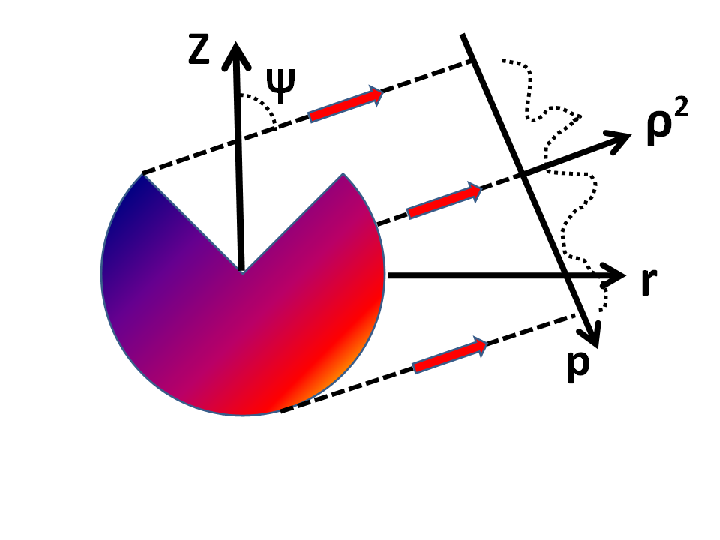}
\caption{Sketch of the coordinate transformation between the original cylindric coordinates (r,z) to the one-dimensional projected coordinate (p) in a  
orthogonal direction to the line of sight defined by the viewing angle $\psi$.  
}
\label{fig13}
\end{figure}

\begin{figure}
\includegraphics[angle=270, scale=0.65]{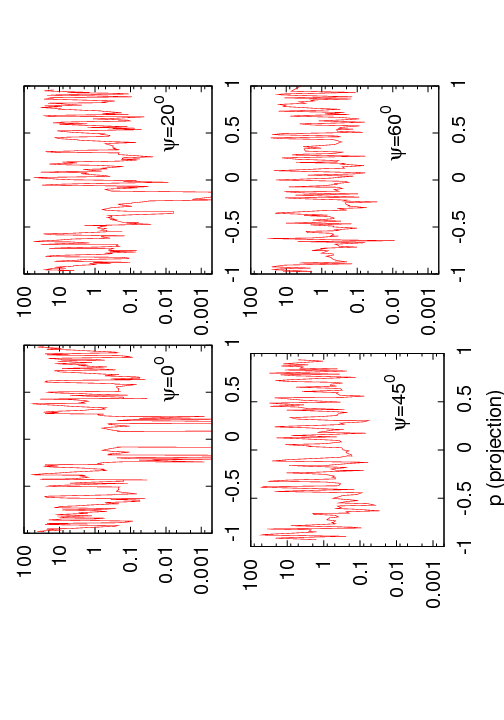}
\caption{Projection of Model E in Table 2 in a orthogonal line to the line of 
sight for several viewing angles $\psi=0^0, 20^0,  45^0$~and $60^0$. The normalized magnitude $\rho^2$~is depicted as a function of coordinate $p$~in the 
projection line.
}
\label{fig14}
\end{figure}

\begin{figure}
\plotone{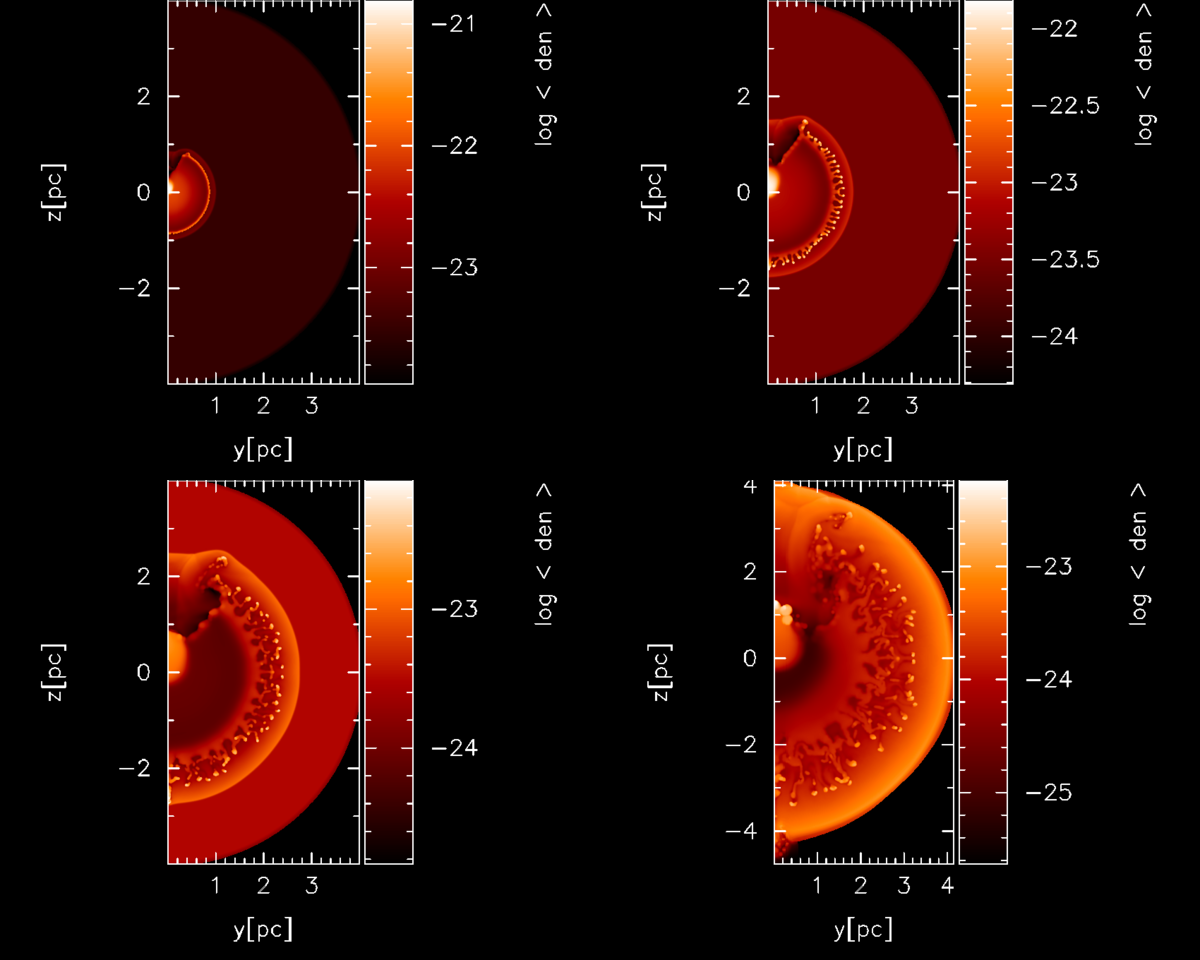}
\caption{ Snapshots summarizing the evolution of the density during
the Sedov-Taylor self-similar phase of the
SNR corresponding to model F of Table 2. 
Adimensional times are t'=0.41, t'=0.86, t'=1.76 and t'=3.9.
}
\label{fig15}
\end{figure}

\begin{figure}
%\plotone{f16.png}
\includegraphics[angle=270, scale=0.65]{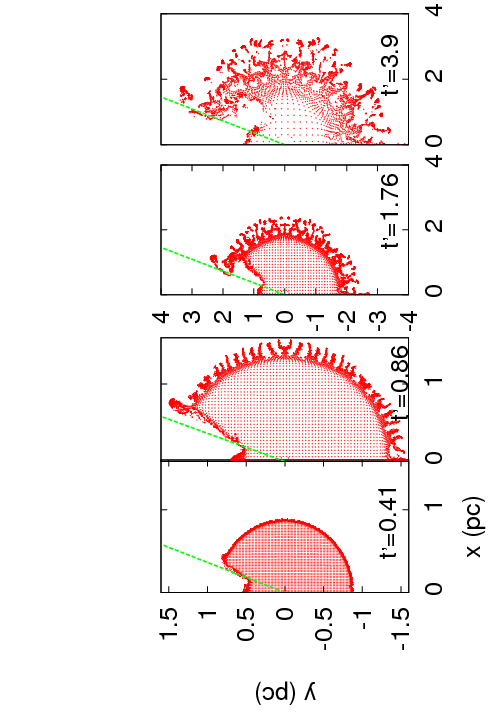}
\caption{Same as in Figure 12 but for model F in Table 2. 
}
\label{fig16}
\end{figure}

\begin{figure}
\includegraphics[angle=270, scale=0.65]{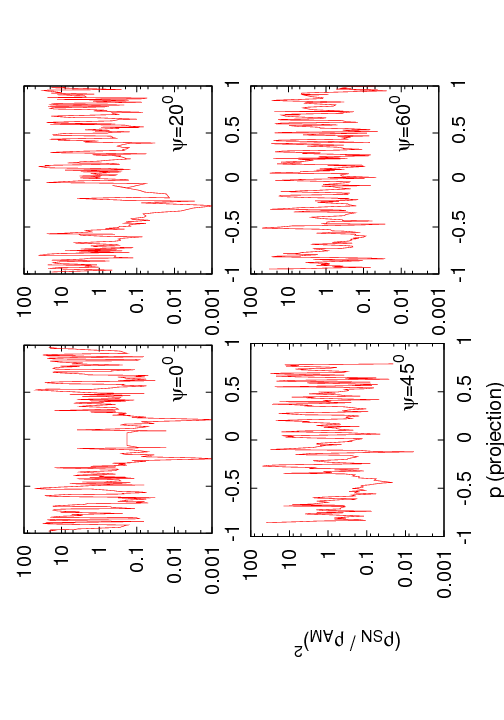}
\caption{Same as in Figure 14 but for model F in Table 2. 
}
\label{fig17}
\end{figure}

\subsection{SNR from an initial ejecta with an empty hole}

We have first considered the situation in which the cone carved in the expanding supernova shell is
initially devoid of any matter, model E in Table 2.  
 This represents a limiting case for those models where 
 the density in the hole would be much smaller than that of the AM at t$_{sim}$=0 yr, being useful to compare to the spherical
ejecta simulation described in the previous section and to the filled hole case described in the next section. Such hypothesis could adequately represent 
scenarios where the stripped matter allocated inside the hole is very small as 
it could be for instance the case of Tycho remnant where the total amount of 
stripped matter 
was estimated to be less than $8~10^{-3}$~M$_{\sun}$ by \cite{lu11}. Several
snapshots of the evolution of the remnant at different times are shown in Figure 10. Shortly after the interaction with
the AM starts the hole begins to be replenished with AM material to finally 
be completely filled in a time $\Delta t'\simeq\frac{R_0}{\bar
v_{SN}~T}$, where $R_0$~is the size of the ejecta at the beginning of the simulation. For model E in Table 2 that time is $\Delta t'=0.12$.  
In the comoving frame moving at the average ejecta velocity 
the initially radial streamlines of AM diverts when they reach the edge of the
hole at $\theta\simeq 20^0$. As the diffracted streamlines gain lateral momentum, they
converge and compress in the neighborhood of the symmetry axis. Eventually, a steady state is reached in which the
forward shock in this region lags behind the rest of the blast wave, distorting its spherical symmetry.  Figure 11 (upper rows) shows
the velocity field in the upper quadrant viewed from the stationary center of our simulation space. At t'$_{\mathrm sim}=0.11$, the hole
is almost filled with AM material, which is compressed toward the hole axis. At t'$_{\mathrm sim}$=0.26, the velocity field in the hole
has already aligned with the symmetry axis and the flow approaches a steady state. As time goes on, the RT instability
develops, as shown in the second snapshot of Figure 10. At the outer edge of the hole the instability growth
is particularly strong owing to the peculiarities of the hydrodyminamic flux in that region. Numerical and laboratory experiments with laser-produced plasma  
\cite{kan01} indicate that hydrodynamic instabilities and vortex like structures form close to the symmetry axis when a supersonic flow of low density material goes through a stationary sphere made of higher density material. Moreover, in our case  
the presence of the hole also breaks the symmetry of the flow.     
 As a result, the RT
fingers at the edge of the hole grow stronger and project inside the hole volume at late times. This intrusion is illustrated in Figure
12, where we have overlayed a dashed line at $\theta=20^0$~for reference. Despite the strongly supersonic nature
of the flow, this simulation suggests that the hole in the ejecta structure caused by the presence of the secondary star
at early times could close or shrink appreciably over timescales of several hundred years due to the RT instability.

In Figure 8 there is shown the location of the most distant particle of the ejecta a a function of the azimuthal angle. The profile can be compared to the that obtained in section 4.1
assuming spherical symmetry. As expected, the largest differences are found for $\theta\leq 20^0$, but minor
differences are also found at other regions in the upper quadrant where $\theta\le 90^0$~ 
   (for example around
$\theta= 50^0$) while differences in the lower quadrant with $90^0 < \theta \le 180^0$~are much smaller. This suggests that the initial
asymmetry caused by the hole is affecting the hydrodynamic behavior of the remnant on a larger scale. In Figure 9
(center), we show the PWS of the CD radius in the upper and lower quadrants of this model.  At low wave numbers, the PWS
in the upper quadrant  is flatter than in the lower one and the slope of the linear fit is a little steeper. These
results, although qualitative, suggest that the presence of a hole in the SN ejecta could be inferred by studying the
geometrical properties of the CD in real SNRs. 

The dependence of the X-ray emissivity, ($\propto\rho^2$), of the shocked ejecta with the line of sight is quantitatively outlined in Figures 13 and 14. The procedure to make the projection of the remnant onto an orthogonal line to the line of sight  is sketched in Figure 13. For a given projecting line, with angle of sight $\psi$, the cylindric coordinates (r,z) and 
density of the closest ejecta particle to 
the observer is recorded. The coordinates of these closests-to-observer 
particles  with angle $\psi$~were projected into an orthogonal line to the line of sight (line $p$~in Figure 13) and magnitude $\rho^2$~was reresented as a function of the unidimensional normalized variable $p$.
The result of the projection for viewing angles $\psi=0^0, 20^0, 45^0, 60^0$~is shown in Figure 14. For a viewing angle of $\psi=0^0$~the observer is directly 
looking down into the hole. In this case the density profile around the origin 
is symmetric and there is a large density constrast between the hole region and its neighborhoods.  As the viewing angle rises the density contrast goes down 
and the density gap moves to negative coordinates in the projection line. 
For viewing angles $\psi \le 60^0$~the fingerprint of the $\theta_H=40^0$~wide hole has already become very weak.    
Thus, on pure geometrical basis and taking $\psi\simeq 45^0$~as an (probably optimistic) upper 
limit to the viewing angle we would expect that $\simeq 25\%$~($\psi=45^0$) to $\simeq 10\%$~($\psi=20^0$) of Type Ia SNR could display inhomogeneities  
in the X-ray emission caused by the hidden hole. As the aperture of the  
hole primarly depends on the distance between the white dwarf and the secondary at the moment of the explosion, which  
 is in turn related with the precise nature of the companion star, the detection of the hole could bring information about that point.

\subsection{SNR from an initial ejecta with a hole loaded with stripped material}

Several hours after the SN Ia explosion, the expanding ejecta is almost spherical except in a cone-like region with its
apex located at the position of the secondary star at the moment of the explosion. This conical region is not empty, but
loaded with H and He-rich material stripped from the companion. In order to model the effect of this material in the
long term evolution of the SNR, we have included the basic features of the stripped gas in a $20^0$~ wedge around the
symmetry axis.  According to Figure 4, the stripped material in the hole region is moving homologously with a
characteristic $\Delta M/\Delta v$~profile which favors the low velocity tail of the distribution. To incorporate the
stripped mass, a $20^0$~wide slice was removed from the SN ejecta. Then we took the same number of particles and
assigned them solar composition and a velocity profile matching the one shown in the left panel of Figure 4.  
  The
mass of each particle was then modified to approximately follow the $\Delta M/\Delta v$~distribution given in the right
panel of Figure 4, with the only constraint that the total amount of stripped material inside the hole equal the
$\simeq 0.035 M_\sun$ obtained in our detailed simulations of the stripping process.

Several snapshots showing  the evolution of the SNR are shown in Figure 15. On the whole, the SNR evolution looks very
similar to the empty hole case 
explored in the previous section. Nevertheless, the addition of the low-velocity material stripped from the secondary introduces 
a few significant differences. For example, the flow inside the hole gets aligned faster with 
the symmetry axis, as can be seen by comparing the upper and lower rows in Figures 11.  
   At the outer edge of the hole, there is a larger stirring 
effect of the ejecta material as it interacts with the low density but 
high velocity component of the stripped matter. As shown in Figure 15, this leads to a stronger 
development of the RT instability in that region. At t'=1.76 the RT finger 
around $\theta\simeq 25^0$~is bigger than in the the empty hole case, and its outer edge is 
close to the forward shock. The high density but 
low velocity component of the stripped material does not have a major impact on the SNR geometry, partially due to the
damping introduced by the artificial viscosity in our SPH simulations, 
which suppresses the growth of instabilities close to the center. To follow the hydrodynamical evolution of the
innermost region of the hole with sufficient detail, a much higher resolution 
study should be conducted. As in the previous section, the strong development 
of the RT instability at the edge of the hole leads to the intrusion of some 
supernova ejecta into the hole (see Fig 16), but to a lesser extent. The final fate of the hole is not clear. Although
the hole may never close completely, our simulations do not 
exclude its partial closure by hydrodynamic instabilities at late times. In any case, the simulations presented here suggest that the hole will 
remain open during several centuries, distorting the geometry ofthe CD in historical Type Ia SNRs.  

The radius of the CD is at t'=3.9 is shown in Figure 8. The profile is similar to the empty hole case, with minor
differences. As discussed above, the hole seems to have closed more in the ejecta profile without stripped mass than in
the one including it. The radial amplitude of the CD at $\theta=25^0$~is bigger owing to the larger vertical
excursion of the RT mushroom in that region. These peculiarities are also present in the PWS of the angular distribution
of radial amplitude of the CD, as shown on the right of Figure 9. The PWS in the upper quadrant is similar to that of
the empty hole case but the features are more pronounced and the slope of the linear fit steeper. The PWS in the lower
quadrant was similar to that of spherical and empty hole calculations, as expected. As in the precedent section we have projected the supernova ejecta onto a
 line orthogonal to the line of sight (Figures 13 and 17) to estimate limiting  viewing angles to detect the hole. The results were similar to that of the 
empty hole case.

\section{Conclusions}

The imprint of the secondary star in the long term evolution of Type Ia SNRs has been studied by means of numerical
hydrodynamic simulations. We simulated first the interaction between the spherically symmetric SN ejecta and the nearby
companion star in the standard SD scenario for Type Ia SNe. We used the results of these simulations to set the initial
conditions for the later phases of the SNR evolution. Our calculations were carried out using an axisymmetric SPH code,
which makes it easy to keep track of the several components of the fluid, namely the supernova material, the AM and the
material stripped from the envelope of the companion star during the initial interaction.

The results of the first phase were in basic agreement with those of \citet{mbf00}, \citet{ser05}, and \citet{pak08} for
the chosen initial configuration of the binary system at the moment of supernova explosion (summarized in Table 1). Of
special relevance for our study were the angular amplitude of the hole opened in the supernova debris by the shielding
effect of the companion, sun-like star, the amount and velocity profile of the stripped material and its distribution in
velocity space in the hole region.

Once the initial interaction ended, an homologous transformation was applied to set the initial conditions for the
second phase and a large volume of AM was incorporated to the computational domain. The ratio between the volumes
encompassed by the AM and the supernova debris at t=0 yr was $\simeq 6000$. Such a huge value highlights the difficulty
to carry out a full three dimensional study of the phenomena with sufficient resolution. The existence of a symmetry line makes it possible to   
handle the evolution of the remnant using an  axisymmetric hidrocode so that the achieved resolution was enough to grasp the main features of the 
of the interaction of the SN ejecta with their surroundings.
 Three calculations were carried out, assuming complete spherical
symmetry in the ejecta, an empty $\theta_H=40^0$~hole due to the presence of the companion, and a similar hole filled with
stripped material from the secondary star, Table 2.

Our simulations show that the hole carved in the ejecta affects the long term evolution of the SNR. We have seen that
hydrodynamic instabilities at the edge of the hole trigger the intrusion of material toward the symmetry axis, especially
in the empty hole model. When the hole is filled with material from the companion star, this intrusion was somewhat
suppressed.  We conclude that the conical hole will remain open during a long time, probably longer than the typical age
of historical SNRs, but its closure over longer timescales is not ruled out. 

We have characterized the influence of the
hole on the geometrical properties of the contact discontinuity separating the supernova material from the AM. Within
the limitations of our study, we showed that the CD has different geometrical properties in the hemisphere that contains
the hole. The hole seems to influence the dynamics of the SNR over angular scales that are larger than its size. The
Fourier analysis of the angular distribution of the radial fluctuation of the CD suggests that the power spectrum of the
CD surface should have a steeper exponent close to the hole than away from it, but this
statement should be confirmed by a full 3D study with sufficient resolution. 

 Another point of interest is that
hydrodynamic instabilities at the outer edge of the hole develop faster than the average growth rate in the RT unstable
region of the SNR, and can come close to the forward shock. The extension 
of the RT instability layer has been addressed in several works, \cite{blo01}, \cite{wan02} where two mechanisms were identified as agents which could significantly enhance 
the penetration of the RT fingers. One is the reduction of the effective 
adiabatic $\gamma$~value used in the EOS due to particle acceleration at 
the forward shock, \cite{blo01}. In this case the larger extent of the region subceptible to the RT instability would be more or less homogeneous, thus affecting the whole unstable shell. The second mechanism invokes the existence of 
isolated dense ejecta clumps in core-collapse SNR to explain the observed protusions in the Vela supernova remnant, \cite{wan02} and would be much more localized than the first one. According to our calculations the existence of a void in
the supernova ejecta, either filled or not with stripped matter from the former 
companion star, may also leads to a larger development of the RT fingers in the outer edge of the hole region.  
 This suggests that, if observed,
the existence of isolated RT structures with anomalous size in putative Type Ia SNR could be 
an indication supporting the SD escenario. 

Nevertheless the limited size of hole would make its detection difficult.  
According to the geometrical analysis given in Sections 4.2 and 4.3 the chance 
to detect the hole is small for viewing angles (with respect to the symmetry axis) $\psi> 30-40^0$~and completely negligible for $\psi>60^0$. For a random SNR distribution that means that only 
one among five or six SNR arising from Type Ia SN explosions which hosts a $\simeq 40^0$~hole may be detected this way. Such probability is lower/higher for 
smaller/wider holes, which in turn depends on the particular features of the 
binary system
where the explosion of the white dwarf took place.    

The novel calculations described in this manuscript confirm that multidimensional hydrodynamics can be safely applied
to the study of the long term evolution of SNR after the explosion of a SN Ia in the single degenerate paradigma. Given that simulations were much time 
consuming we 
have focussed on a particular combination of parameters of the binary 
system hosting the explosion and used an uniform ambient medium to  
explore the 
consequences on the long term evolution of the remnant. Therefore our 
analysis does not apply to a particular SNR although we expect that  
the gross features of the phenomena were captured by our simulations.  

\acknowledgements

We thank the anonymous referee for the constructive critical comments that much improved the presentation of the manuscript. This work was funded by Spanish MICINN grants 
AYA2010-15685, AYA2008-04211-C02-01 and also supported  
by DURSI of the Generalitat de Catalunya. CB acknowledges support from the Benoziyo Center for Astrophysics and an EU FP7 Marie Curie IRG Fellowship. The rendered SPH plots were made using the freely available {\it SPLASH} code written by Daniel Price (2007).  

%\clearpage

%%
%% End of file `sample.tex'm

\end{document}